\documentclass[reprint,superscriptaddress,amsmath,amssymb,aps,prl]{revtex4-2}

\usepackage{graphicx}
\usepackage{dcolumn}
\usepackage{bm}
\usepackage{hyperref}
\usepackage{color}
\definecolor{blue2}{rgb}{0.12,0.15,0.52}
\hypersetup{colorlinks=true,urlcolor=blue2,citecolor=blue2,linkcolor=blue2}

\newcommand{\tr}{\mathrm{tr}\,}

\begin{document}

\title{Geometric formulation of generalized root-$T\bar{T}$ deformations }

\affiliation{Center for Theoretical Physics and College of Physics, Jilin University, 
Changchun 130012, China}

\author{H. Babaei-Aghbolagh}
\email{h.babaei@uma.ac.ir}
\affiliation{Department of Physics, University of Mohaghegh Ardabili, P.O. Box 179, Ardabil, Iran}

\author{Song He}
\email{hesong@jlu.edu.cn}
\affiliation{Center for Theoretical Physics and College of Physics, Jilin University, 
Changchun 130012, China}

\affiliation{School of Physical Science and Technology, Ningbo University, Ningbo, 315211, China}

\affiliation{Max Planck Institute for Gravitational Physics (Albert Einstein Institute),
Am M\"uhlenberg 1, 14476 Golm, Germany}

\author{Tommaso Morone}
\email{tommaso.morone@unito.it}
\affiliation{Dipartimento di Fisica, Università di Torino, and INFN Sezione di Torino, Via P. Giuria 1, 10125,
Torino, Italy}

\author{Hao Ouyang}
\email{haoouyang@jlu.edu.cn}
\affiliation{Center for Theoretical Physics and College of Physics, Jilin University, 
Changchun 130012, China}

\author{Roberto Tateo}
\email{roberto.tateo@unito.it}
\affiliation{Dipartimento di Fisica, Università di Torino, and INFN Sezione di Torino, Via P. Giuria 1, 10125,
Torino, Italy}

\begin{abstract}
We develop a generic geometric formalism that incorporates both $T\bar{T}$-like and root-$T\bar{T}$-like deformations in arbitrary dimensions. This framework applies to a wide family of stress-energy tensor perturbations and encompasses various well-known field theories. Building upon the recently proposed correspondence between Ricci-based gravity and $T\bar{T}$-like deformations, we further extend this duality to include root-$T\bar{T}$-like perturbations. This refinement extends the potential applications of our approach and contributes to a deeper exploration of the interplay between stress tensor perturbations and gravitational dynamics.
Among the various original outcomes detailed in this article, we have also obtained a deformation of the flat Jackiw-Teitelboim gravity action.

\end{abstract}

\maketitle


\section{Introduction}
Recent studies concerning deformations of classical and quantum field theories have revealed rich connections between geometry and field dynamics. A prime example is that of $T\bar{T}$ deformations \cite{Cavaglia:2016oda, Smirnov:2016lqw} of two-dimensional theories, driven by the irrelevant composite operator \cite{Zamolodchikov:2004ce}
 \begin{equation}
 O_{T\bar T}=-\mathrm{det}\left(T^{\mu}{}_\nu\right)=\frac12\left(T^{\mu\nu}T_{\mu\nu}-T^{\mu}{}_{\mu}T^{\nu}{}_{\nu}\right).
 \end{equation}
Despite being irrelevant, two-dimensional $T\bar{T}$ deformations remain well-controlled and even solvable at the quantum level. In the deformed theory, various quantities can be computed exactly from their counterparts in the original model. These include the finite-volume spectrum, the S-matrix \cite{Smirnov:2016lqw,Cavaglia:2016oda}, the  classical Lagrangian \cite{Bonelli:2018kik,Conti:2018jho,Conti:2018tca}, and the torus partition function \cite{Cardy:2018sdv,Datta:2018thy,Aharony:2018bad, He:2020cxp}.
$T\bar{T}$ deformations connect with different topics in theoretical physics, such as string theory \cite{Baggio:2018gct,Dei:2018jyj,Chakraborty:2019mdf,Callebaut:2019omt}, holography \cite{McGough:2016lol,Kraus:2018xrn,Cottrell:2018skz,Taylor:2018xcy,Hartman:2018tkw,Caputa:2019pam,Guica:2019nzm,Gross:2019ach,Li:2020pwa,He:2023hoj,He:2023knl,Ebert:2024fpc}, and quantum gravity \cite{Dubovsky:2017cnj,Dubovsky:2018bmo,Tolley:2019nmm,Iliesiu:2020zld,Okumura:2020dzb,Ebert:2022ehb}.  We refer the reader to \cite{Jiang:2019epa} for a pedagogical review on the subject.

Furthermore, the $T\bar T$ deformation lends itself to a number of geometric interpretations.
It was proposed in \cite{Cardy:2018sdv} that $T\bar T$-perturbing a theory is equivalent to coupling the original theory to a random geometry.
$T\bar T$ deformations can also be interpreted as coupling the original theory to a flat space Jackiw-Teitelboim-like gravity \cite{Dubovsky:2017cnj,Dubovsky:2018bmo},
or equivalently, a topological gravity \cite{Tolley:2019nmm, Caputa:2020lpa, Dubovsky:2023lza}.

Another interesting deformation of two-dimensional field theories, driven by the so-called root-$T\bar{T}$ operator  \cite{Conti:2022egv,Babaei-Aghbolagh:2022leo,Ferko:2022cix}
\begin{equation}
 R=\sqrt{\frac12 T^{\mu\nu}T_{\mu\nu}-\frac14 T^{\mu}{}_{\mu}T^{\nu}{}_{\nu}},
 \end{equation}
has recently attracted growing attention. While its quantum-mechanical definition remains uncertain, the root-$T\bar{T}$ perturbation displays some surprising properties at the classical level. Notably, it commutes with the $T\bar{T}$, allowing for their simultaneous activation, and for some integrable field theories, it preserves classical integrability \cite{Borsato:2022tmu}.
The relation between root-$T\bar{T}$ deformed conformal field theories and ultra-relativistic (BMS$_{3}$)  field theories was discussed in \cite{Rodriguez:2021tcz}. Finally, the connection between the root-$T\bar{T}$ deformation and the modified boundary conditions in the holographic dictionary was studied in \cite{Ebert:2023tih}. These results were later employed to explore the modular properties of deformed holographic conformal field theories in \cite{Tempo:2022ndz, Tian:2024vln}.

In higher space-time dimensions, stress-energy tensor perturbations give rise to many interesting field theory models \cite{Cardy:2018sdv,Bonelli:2018kik,Conti:2022egv}. Extensive research has focused on $T\bar{T}$-like and root-$T\bar{T}$-like deformations of four-dimensional Maxwell's theory, exploring the relationship between electromagnetic duality invariance and stress tensor deformations \cite{Babaei-Aghbolagh:2020kjg,Babaei-Aghbolagh:2022itg,Ferko:2023wyi}. The massive gravity formulation of duality-invariant non-linear electrodynamics was studied in \cite{Floss:2023nod} and, in three dimensions, it was shown that Born-Infeld theory displays a classical $T\bar{T}$-like flow, connected to free Maxwell theory \cite{Ferko:2023sps}. Furthermore, recent studies have explored nonlinear chiral two-form gauge theories in six dimensions as $T\bar{T}$-like deformations \cite{Ferko:2024zth}.

This paper introduces a generic geometric approach to encompass a broader class of stress-energy tensor perturbations. We show that a two-dimensional theory deformed by both $T\bar{T}$ and root-$T\bar{T}$ operators is dynamically equivalent to the undeformed theory coupled to a novel gravity action, at least at a classical level. We further generalize the geometric formulation to accommodate various deformations in higher dimensions. While prior studies have investigated geometric formulations of $T\bar{T}$-like deformations in higher dimensions within the metric approach \cite{Conti:2022egv}, our formulation is based on the description in terms of eigenvalues of the product of the vielbein. This approach allows us to study stress-energy tensor-related flows within a simple and elegant setup.

The recent work \cite{Morone:2024ffm} has emphasized that a $T\bar{T}$-type deformed matter action coupled with the standard Einstein-Hilbert action is equivalent to an undeformed matter theory coupled with a Ricci-based gravity theory \cite{Olmo:2022rhf}. Adopting this perspective, we incorporate this logic into our geometric formulations and introduce Ricci-based gravity actions linked with root-$T\bar{T}$-like deformations.
We develop a unified framework for 
\(T\bar{T}\) and root-\(T\bar{T}\) perturbations in field theories across various space-time dimensions, 
which may extend the class of exact-solvability preserving deformations and deepen our understanding of the fundamental principles of quantum gravity and string theory.


\section{Unified Geometric Formulation of $T\bar{T}$ and Root-$T\bar{T}$ Deformations in $d=2$}\label{sec:2d}

We denote by $S_0[\phi,e^a_\mu]$ an arbitrary undeformed action, where $\phi$ indicates a generic collection of matter fields and $e^a_\nu$ denotes an auxiliary dynamical zweibein.
The associated auxiliary metric is $g_{\mu\nu}=\eta_{ab}e^a_\mu e^b_\nu$.
We couple the auxiliary zweibein to a second zweibein $f^a_\mu$, and the metric tensor $h_{\mu\nu}=\eta_{ab}f^a_\mu f^b_\nu$ associated to $f^a_\mu$ will eventually emerge as the metric of the manifold on which the deformed theory lives. It is convenient to define two Lorentz invariant variables
\begin{equation}
\begin{split}
    y_1&=\mathrm{tr}(e^{-1}f)=f^a_\mu e_a^\mu ,\\~~~
    y_2&=\mathrm{tr}[(e^{-1}f)^2]=f^a_\mu e_b^\mu f^b_\nu e_a^\nu.    
\end{split}
\end{equation}
We now show that the combination of $T\bar{T}$ and root-$T\bar{T}$ deformations can be generated from the action:
\begin{equation}\label{2drt}
    S_{\gamma,\lambda}[\phi,e^a_\mu,f^a_\mu]=S_0[\phi,e^a_\mu]+S_{\mathrm{grav}}[e^a_\mu,f^a_\mu],
\end{equation}
where gravity action $S_{\mathrm{grav}}$ is
\begin{equation}\label{se}
\begin{split}
  &S_{\mathrm{grav}}[e^a_\mu,f^a_\mu]
 = \frac{1}{2\lambda}\int d^2x \det e
 \\
&\times\left(
 2+y_1^2-y_2-2y_1 \cosh \frac{\gamma }{2}
 +2\sqrt{2y_2-y_1^2}
 \sinh \frac{\gamma }{2}
 \right).    
\end{split}
\end{equation}
The parameters $\lambda$ and $\gamma$ represent the $T\bar T$ and the root-$T\bar T$ perturbing parameters, respectively.
When $\gamma=0$, $S_{\mathrm{grav}}$ reduces to the topological gravity action associated to the $T\bar{T}$ deformation \cite{Tolley:2019nmm}:
\begin{equation}
S_{\mathrm{grav}}[e_{\mu}^{a},f_{\mu}^{a}]=\frac{1}{2\lambda}\int d^{2}x\epsilon^{\mu\nu}\epsilon_{ab}(e_{\mu}^{a}-f_{\mu}^{a})(e_{\nu}^{b}-f_{\nu}^{b}),
\end{equation}
where $\epsilon$ is the Levi-Civita symbol.
Our analysis will be carried out using the Euclidean signature, and the generalization to the Lorentzian signature is straightforward.

The deformed action can be obtained by extremizing (\ref{2drt}) with respect to the auxiliary zweibein $e^a_\mu$: performing the variation of (\ref{2drt}) with respect to $e^a_\mu$, we have
\begin{equation}\label{on-shell-e}
    \det e\, (T^{[0]})^\mu_{~\nu}\equiv \frac{\delta S_{0}}{\delta e^a_\mu}e^a_\nu=-\frac{\delta S_{\mathrm{grav}}}{\delta e^a_\mu}e^a_\nu,
\end{equation}
where $(T^{[0]})^\mu_{~\nu}$ is the stress-energy tensor of the undeformed theory, computed with respect to $e^a_\mu$. We denote the solution of the equation of motion by ${e^*}^a_\mu$. Note that \eqref{on-shell-e} may admit multiple solutions ${e^*}^a_\mu$ related to the choice of the branch for the square root of the root-$T\bar{T}$ operator: in this work, we ignore such branch ambiguities. However, in the quantum theory, we expect one should sum over contributions from all branches in the path integral. The deformed field theory is obtained substituting ${e^*}^a_\mu$ back into (\ref{2drt}):
\begin{equation}
S_{\mathrm{deformed}}[\phi,f^a_\mu]=S_{\gamma,\lambda}[\phi,{e^*}^a_\mu,f^a_\mu].
\end{equation}
The stress-energy tensor of the deformed theory can be computed as
\begin{equation}
    T^\mu_{~\nu}\equiv \frac{1}{\det f}\frac{\delta S_{\gamma,\lambda}}{\delta f^a_\mu} f^a_\nu=\frac{1}{\det f}\frac{\delta S_{\mathrm{grav}}}{\delta f^a_\mu}f^a_\nu\Big|_{e=e^*},
\end{equation}
where we have used the on-shell condition \eqref{on-shell-e} for $e^a_\mu$, so that $S_{\lambda}[\phi,{e^*}^a_\mu,f^a_\mu]$ explicitly depends on $f^a_\mu$ alone.
To simplify notation, we will not distinguish between $e^a_\nu$ and its on-shell value ${e^*}^a_\mu$, unless necessary.
One can verify that the total action (\ref{2drt}) obeys the following flow equations:
\begin{align}
    \label{flow1}\frac{\partial S_{\gamma,\lambda}}{\partial \lambda}&=-\int d^2x \det f \det(T^\nu_{~\mu}),\\
    \label{flow2}\frac{\partial S_{\gamma,\lambda}}{\partial \gamma}&=\int d^2x \det f \sqrt{\frac{1}{2}T^\mu_\nu T^\nu_\mu-\frac{1}{4}(T^\nu_\nu)^2}.
\end{align}
Therefore, the action (\ref{2drt}) provides a geometric description of the combined $T\bar{T}$ and root-$T\bar{T}$ deformations. Since $S_{\gamma,\lambda}$ is defined as independent of the flow path, and since the operators do not have explicit $\lambda$ and $\gamma$ dependence, the two types of deformations commute with each other. 

As discussed in Section 2.3 of \cite{Tolley:2019nmm}, one can translate the vielbein formulation to the metric formulation by choosing a gauge such that $e^{-1}f=\sqrt{g^{-1}h}$ by using the local Lorentz transformations of $e$ and $f$, 
where we have omitted indices to simplify the notation.
The validity of the flow equations \eqref{flow1} and \eqref{flow2} can also be verified in the metric formulation, and the details are shown in the Supplemental Material.

We now illustrate our methodology, starting from  the simple undeformed action of a free scalar:
\begin{equation}
    S_0[\phi,e^a_\mu]=
    \int d^2x \det e
   \left(\frac{1}{2}\eta^{ab}e_a^\mu e_b^\nu \partial_\mu \phi \partial_\nu \phi \right).
\end{equation}
The solution of the equation of motion for $e^a_\nu$ is
\begin{equation}
\begin{split}
 &{e^*}^a_\mu
=\frac{1}{2} e^{\frac{\pm\gamma }{2}} \left(\frac{1}{\sqrt{1-4 \lambda  e^{\pm\gamma } X}}+1\right) f^a_\mu \\
&
\mp \left(
\frac{  \sinh \frac{\gamma }{2}\pm 2 \lambda  e^{\pm\frac{\gamma }{2}} X}{ X \sqrt{1-4 \lambda  e^{\pm\gamma } X}}+\frac{ \sinh \frac{\gamma }{2}}{2 X}
\right)
\eta^{ab}f_b^\nu \partial_\nu \phi \partial_\mu \phi  ,   
\end{split}
\end{equation}
where $X=\frac{1}{2}\eta^{ab}f_a^\mu f_b^\nu \partial_\mu \phi \partial_\nu \phi$.
Substituting the solution ${e^*}^a_\mu$ into the action, we get,
\begin{equation}
 S_{\gamma,\lambda}[\phi,{e^*}^a_\nu,f^a_\nu]  
 =\int d^2x
 \frac{1- \sqrt{1-4 e^{\pm\gamma }\lambda  X}}{2 \lambda },
\end{equation}
which reproduces the result obtained in \cite{Ferko:2022cix}.
\section{Uplift to higher dimensions}\label{sec:gen}
In this section, we uplift the geometric description to a family of deformations induced by functionals of the stress-energy tensor in higher dimensions.
In $d$ space-time dimensions, we consider the following general form for the gravity action:
\begin{equation}
\label{eqB}
S_{\mathrm{grav}}[e^a_\mu,f^a_\mu]
 = \int d^d x \det e B(e^{-1}f),
\end{equation}
where $B$ is a Lorentz invariant function of $(e^{-1}f)^{\mu}_{~ \nu}$. Therefore, $B$ depends only on the Lorentz invariant variables
$y_n=\mathrm{tr}[(e^{-1}f)^n] $, $n=1,...,d$.
For $n>d$, the $y_n$ are not independent quantities.
Since one can express $B$ in terms of the variables $y_n$, the stress-energy tensor can be computed as
\begin{equation}\label{set17}
T\equiv \frac{1}{\det f}\frac{\delta S_{\mathrm{grav}}}{\delta f}f
=\sum_{n=1}^d \frac{n}{\det (e^{-1} f)} (e^{-1} f)^n \partial_{y_n} B,
\end{equation}
where $T$ denotes the matrix $T^\mu_{~\nu}$. To construct higher-dimensional deforming operators, we need to compute Lorentz invariant functionals of the stress-energy tensor \eqref{set17}. Although we can express each invariant $\mathrm{tr}(T^k)$ in terms of the $y$-variables, this approach is quite inefficient in arbitrary dimensions, since there is no simple general formula for $y_n$ when $n>d$.

However, assuming $e^{-1}f$ can be diagonalized by means of some matrix $U$ as $e^{-1}f= U\, \mathrm{diag}(\alpha_1,...,\alpha_d)\,U^{-1}$, the function $B$ can be expressed in terms of the eigenvalues $\alpha_i$, and each $y_n$ reduces to a power sum symmetric polynomial of the $\alpha_i$.
For this reason, working with the eigenvalues of $e^{-1}f$ proves to be a far more convenient strategy.
The stress-energy tensor can be expressed as
\begin{equation}\label{stress-tensor-18}
   T= (\prod_{k=1}^d \alpha_k)^{-1}U\,\mathrm{diag}(\alpha_1\partial_{\alpha_1}B,...,\alpha_d\partial_{\alpha_d}B) \, U^{-1}, 
\end{equation}
and
\begin{equation}
\mathrm{tr}(T^k)=(\prod_{j=1}^d \alpha_j)^{-k}
\sum_{i=1}^d(\alpha_i\partial_{\alpha_i}B)^k.
\end{equation}
Expressing $y_1$ and $y_2$ in terms of eigenvalues of $e^{-1}f$, the two-dimensional gravity action (\ref{se}) can be significantly simplified:
\begin{equation}\label{s2sim}
  S_{\mathrm{grav}}[e^a_\mu,f^a_\mu]
 = \frac{1}{\lambda}\int d^2x \det e
(\alpha_1-e^{\frac{\gamma}{2}})(\alpha_2-e^{-\frac{\gamma}{2}}).
\end{equation}
Note that $S_{\mathrm{grav}}$ is not a symmetric function of the eigenvalues because of the non-analyticity of the root-$T\bar{T}$ operator. Exchanging two eigenvalues is equivalent to crossing a branch cut.

Motivated by  the expression (\ref{s2sim}), we propose a generalization in arbitrary $d$ space-time dimensions:
\begin{equation}\label{Bgen}
    B=\frac{1}{\lambda^{\Sigma-1}}   \prod_{k=1}^d (\alpha_k^{p_k}-\beta_k^{p_k})^{1/p_k},
\end{equation}
where $\lambda$ and $\beta_k$ are perturbing parameters, $p_k$ are numbers characterizing the deformation, and  $\Sigma=\sum_{k=1}^dp_k^{-1}$.
 When $d=2$ and $p_k=1$, the action (\ref{Bgen}) reduces to (\ref{s2sim}) if we identify $\beta_1=e^{\frac{\gamma}{2}}$ and $\beta_2=e^{-\frac{\gamma}{2}}$.
We will show that the parameters $\lambda$ and $\log\beta_k$ emerge as higher-dimensional analogs of the two-dimensional $T\bar{T}$ and root-${T\bar{T}}$ deformation parameters, respectively. With the ansatz \eqref{Bgen}, the eigenvalues of the stress-energy tensor can be computed as
\begin{equation}\label{eigent}
    \tau_i=
\alpha_i\partial_{\alpha_i}B \prod_{j=1}^d \alpha^{-1}_j=\frac{\alpha_i^{p_i}}{\alpha_i^{p_i}-\beta_i^{p_i}}B \prod_{j=1}^d \alpha^{-1}_j.
\end{equation}
We also find
\begin{equation}
  \prod_{k=1}^d \tau_k^{1/p_k}=\frac{1}{\lambda^{\Sigma-1}}
   B^{\Sigma-1} (\prod_{k=1}^d \alpha_k)^{1-\Sigma}.
\end{equation}
Therefore, the flow equation for $\lambda$ is
\begin{equation}\label{flowgen}
    \frac{\partial S_{\mathrm{grav}}}{\partial \lambda}
    =-(\Sigma-1)\int d^d x \det f\, \Big(\prod_{k=1}^d \tau_k^{1/p_k}\Big)^{\frac{1}{\Sigma-1}}.   
\end{equation}
The operator on the right-hand side of the equation (\ref{flowgen}) is non-analytic and not symmetric in terms of the stress-energy tensor eigenvalues $\tau_i$.
Particularly, when all $p_k$ are equal to $p$, we obtain a $( \det T)^{\frac{1}{d-p}}$ deformation \cite{Cardy:2018sdv,Bonelli:2018kik}.
When $\Sigma=2$, the deformation is of order $O(T^2)$. Let us now consider the flow equation for the $\beta$-parameters. We have:
\begin{equation} 
\label{B-tau}
   \beta_i\partial_{\beta_i}B-\beta_j\partial_{\beta_j}B=-(\tau_i-\tau_j)\prod_{k=1}^d \alpha_k.
\end{equation}
Equation (\ref{B-tau}) suggests that the flow should be confined to the surface defined by $\prod_{k=1}^d \beta_k=1$. Otherwise, the perturbing operator would explicitly depend on $\lambda$ and $\beta_k$.
The resulting flow equation is:
\begin{equation}\label{flowbeta}
\sum_{k=1}^{d}v^k\frac{\partial S_{\beta,\lambda}}{\partial \log \beta_k}
    =-\int d^d x \det f\,
    \Big(\sum_{k=1}^{d}v^k \tau_k\Big),
\end{equation}
where $v^k$ are constants satisfying $\sum_{k=1}^{d}v^k=1$.
Varying the $\beta$-parameters on the surface $\prod_{k=1}^d \beta_k=1$ leads to non-analytic marginal deformations that commute with the $\prod_{k=1}^d \tau_k^{1/p_k}$ deformation. In two dimensions, the root-${T\bar{T}}  $ operator can be understood as the difference $\tau_1-\tau_2$.  However, explicitly expressing the difference between the $\tau_k$ in terms of $\mathrm{tr} \,(T^j)$ is more difficult in higher dimensions.
Let us now examine the initial conditions of the flow equations.
When integrating out the auxiliary vielbein $e^a_\mu$, one needs the equations of motion of $e^a_\mu$:
\begin{equation}\label{2dT0}
  B(e^{-1}f)\delta^\mu_{\nu}- (e^{-1}f)^\mu_{\alpha}\frac{\partial B}{\partial (e^{-1}f)^\nu_{\alpha}}=
  -\frac{1}{\det e}\frac{\delta S_{0}}{\delta e^a_\mu} e^a_\nu.
\end{equation}
The right-hand side is finite in the limit $\lambda \rightarrow 0$.
Denoting the eigenvalues of $(T^{[0]})^\mu_{~\nu}$ as $\tau^{[0]}_k$, the solution is
\begin{equation}\label{holo}
\alpha_j=\beta_j\left(\lambda(\tau^{[0]}_j)^{-1}\Big(\prod_{k=1}^d \tau^{[0]}_k{}^{1/p_k}\Big)^{\frac{1}{\Sigma-1}}+1\right)^{1/p_j},
\end{equation}
 which implies that $\alpha_k=\beta_k+O(\lambda)$ when $\lambda \rightarrow 0$. When $\beta_k=1$, we have $e^a_\mu\rightarrow f^a_\mu$ and the total action $S_{\beta,\lambda}=S_0[\phi,e^a_\mu]+S_{\mathrm{grav}}[e^a_\mu,f^a_\mu]$ reduces to the original action $S_0[\phi,f^a_\mu]$. Equation (\ref{holo}) can be interpreted as the deformed boundary conditions in holography, formulated in terms of eigenvalue variables. In the Supplemental Material, we reproduce the root-$T\bar T$ deformed boundary conditions proposed in \cite{Ebert:2023tih}.

\section{Examples}
Several deformed field theories can be explored within this framework. A notable example is the ModMax theory \cite{Bandos:2020jsw} and its Born-Infeld-like (MMBI) extension  \cite{Bandos:2020hgy}.
The ModMax theory is a non-linear conformal- and duality-invariant modification of Maxwell's theory. The MMBI extension maintains the duality invariance, and the action satisfies two commuting flow-equations \cite{Conti:2018jho,Babaei-Aghbolagh:2022uij,Ferko:2022iru}:
\begin{align}
   \frac{\partial\mathcal{L}_{\mathrm{MMBI}}}{\partial \tilde\lambda} &=\frac{1}{8}\left(T_{\mu\nu}T^{\mu\nu}-\frac{1}{2}T_{\mu}{}^{\mu}T_{\nu}{}^{\nu}\right),\label{flowlambda}\\    \frac{\partial\mathcal{L}_{\mathrm{MMBI}}}{\partial \tilde\gamma}
    &=\frac{1}{2}\sqrt{T_{\mu\nu}T^{\mu\nu}-\frac{1}{4}T_{\mu}{}^{\mu}T_{\nu}{}^{\nu}}.\label{flowgamma}
\end{align}
In the MMBI theory, the stress-energy tensor admits two degenerate eigenvalues $\tau_1$ and $\tau_2$, each of multiplicity 2. Therefore, the following relations hold:
\begin{align}
\mathrm{tr}(T^2)-\frac{1}{2}(\mathrm{tr}\,T )^2=-4 \sqrt{\det\,T} = -4\,\tau_1\tau_2,\\
\sqrt {\mathrm{tr}(T^2)-\frac{1}{4}(\mathrm{tr}\,T )^2}=\tau_1-\tau_2.
\end{align}
Turning off the irrelevant deformation momentarily, one can notice that the flow \eqref{flowgen} is satisfied in $d=4$ by fixing $p_k=2$ for each $k$, up to rescaling the irrelevant flow parameter. On the other hand, setting $\beta_1=\beta_2=e^{\gamma}=\beta_3^{-1}=\beta_4^{-1}$, the flow  equation (\ref{flowbeta}) can be identified with (\ref{flowgamma}), up to a rescaling of $\gamma$. This shows that the MMBI flows \eqref{flowlambda} and \eqref{flowgamma} can be realized by coupling Maxwell's theory to the gravity action (\ref{Bgen}) with $d=4$ and $p_k=2$. In this case, (\ref{Bgen}) simply reduces to
\begin{equation}\label{mmbigrav}
    S_{\mathrm{grav}}[e^a_\mu,f^a_\mu]
 = \int d^4 x \det e \Big[ \frac{1}{\lambda} \prod_{k=1}^4 (\alpha_k^2-\beta_k^2)^{1/2} \Big].
\end{equation}
Note that the quantities $\alpha_k^2$ represent the eigenvalues of $g^{\mu\rho} h_{\rho\nu}$: if we switch off
the deformation induced by the $\beta_k$'s, the corresponding action can be expressed explicitly in terms of the metrics:
\begin{equation}
    S_{\mathrm{grav}}[h_{\mu\nu},g_{\mu\nu}]
 = \frac{1}{\lambda}\int d^4 x \sqrt{\det (h_{\mu\nu}-g_{\mu\nu})} .
\end{equation}
$T\bar{T}$-like flows of six-dimensional two-form chiral theories were recently studied in \cite{Ferko:2024zth}.
In these models, $T$ admits two degenerate eigenvalues of multiplicity 3 (throughout the flow), implying that our geometric construction can be straightforwardly implemented. Another example is the higher-dimensional generalized Nambu-Goto action of a self-interacting scalar field in $d$ dimensions:
\begin{equation}\label{hdNG}
 S_{\lambda}
 =\int d^d x\left[
 \frac{1- \sqrt{1-2  \lambda(1-\lambda V) \partial^\mu\phi \partial
_\mu\phi }}{ \lambda (1-\lambda V) } - \frac{2V}{1-\lambda V}\right].   
\end{equation}
The action \eqref{hdNG} satisfies the flow equation
(the $V=0$ case has been proven in \cite{Ferko:2024zth})
\begin{equation}\label{flowFerko}
\begin{split}
    \frac{\partial S_{\lambda}}{\partial \lambda}=&\int d^dx  
    \Big( 
    \frac{1}{2d}\mathrm{tr}(T^2)-\frac{1}{d^2} (\mathrm{tr}\,T)^2\\
   & -\frac{d-2}{2 \sqrt{d-1} d^{3/2}}\mathrm{tr}(T)
\sqrt{\mathrm{tr}(T^2)-\frac{1}{d} (\mathrm{tr}T)^2}
    \Big).    
\end{split}
\end{equation}
The stress-energy tensor has a non-degenerate eigenvalue $\tau_1$ and a degenerate eigenvalue $\tau_2$ of multiplicity $d-1$. In terms of eigenvalues, the deforming operator can be written as $-\frac{1}{2}\tau_1 \tau_2$.
Therefore, such deformation can be achieved by setting
$\beta_j=1$, $p_1=1$ and $p_{k>2}=d-1$
in (\ref{flowgen}). 
It was shown in \cite{Ferko:2023sps} that the three-dimensional Born-Infeld theory also satisfies the flow equation (\ref{flowFerko}), and one can show that $T$ has a non-degenerate eigenvalue $\tau_1$ and a degenerate eigenvalue $\tau_2$ of multiplicity 2, allowing for a similar description of the flow.
Finally, alternative geometric formulations can be constructed when the theory's stress-energy tensor has two distinct degenerate eigenvalues, as described in the Supplemental Material.
\section{ Inclusion of dynamical gravity}\label{sec:gra}
In \cite{Morone:2024ffm}, it was pointed out that a $T \bar T$ deformed matter action coupled to the Einstein-Hilbert action is equivalent to an undeformed matter theory coupled to a Ricci-based gravity. Continuing along the same line of thought, we now make the metric $h$ dynamical and include the Einstein-Hilbert term within the first-order Palatini formalism. The total action is:
\begin{equation}\label{metric3}
\begin{split}
 S[h, g,\Gamma,\phi]=&
    \frac{1}{2\kappa}\int d^d x \sqrt{\det h} h^{\mu\nu}R_{\mu\nu}(\Gamma)\\
   & +\int d^d x \sqrt{\det g}B(g^{-1}h)+
    S_0[g,\phi],
\end{split}
\end{equation}
where the Ricci curvature tensor is a functional of the connection
\begin{equation}
R_{\mu\nu}(\Gamma)=\partial_\alpha\Gamma_{\nu\mu}^\alpha-\partial_\nu\Gamma_{\alpha\mu}^\alpha+\Gamma_{\alpha\beta}^\alpha\Gamma_{\nu\mu}^\beta-\Gamma_{\nu\beta}^\alpha\Gamma_{\alpha\mu}^\beta.
\end{equation}
The equations of motion for the connection $\Gamma_{\mu\nu}^{\lambda}$ lead to the compatibility conditions
\begin{equation}    \Gamma_{\mu\nu}^{\lambda}=\frac{1}{2}\left(h^{-1}\right)^{\lambda\alpha}\left(\partial_{\nu}h_{\mu\alpha}+\partial_{\mu}h_{\alpha\nu}-\partial_{\alpha}h_{\mu\nu}\right).
\end{equation}
Integrating out $g$ in the action (\ref{metric3}) we get
\begin{equation}        S[h,\Gamma,\phi]=
    \frac{1}{2\kappa}\int d^d x \sqrt{\det h} h^{\mu\nu}R_{\mu\nu}(\Gamma)
    +
    S_{\mathrm{deformed}}[h,\phi],
\end{equation}
which can be viewed as a deformed matter action $ S_{\mathrm{deformed}}$ coupled to the standard Einstein-Hilbert action.

To obtain the Ricci-based gravity description, one can integrate out $h$ in the action (\ref{metric3}) and obtain,
\begin{align}
 & S[g,\Gamma,\phi]=
   \int d^d x \sqrt{\det g} \mathcal{L}(g^{-1} R)
    +
    S_{0}[g,\phi],\\
& \mathcal{L}=  
\left[ B(g^{-1}h)- \frac{1}{d-2}g^{\mu\alpha}h_{\alpha\nu}\frac{\partial B}{\partial (g^{\mu\beta}h_{\beta\nu})} 
\right] \bigg|_{h=h^*(g)},
\end{align}        
which can be interpreted as an undeformed matter action coupled to a Ricci-based gravity theory $\mathcal{L}(g^{-1} R)$.
This procedure yields a dynamical equivalence between an undeformed matter theory coupled to a Ricci-based gravity and a deformed theory
coupled to standard general relativity. 

It is, however, difficult to obtain an explicit expression for the Lagrangian $\mathcal{L}(g^{-1} R)$ associated with the $B$ function given by (\ref{Bgen}) because the equations of motion of $h$ are in general very complicated. In the Supplemental Material, we derive a flow equation for $\mathcal{L}(g^{-1} R)$:
\begin{equation}\label{flowL}
\begin{split}
  \frac{\partial \mathcal{L}(\rho)}{\partial \lambda}=&-(\Sigma-1)\kappa^{-\frac{\Sigma}{\Sigma-1}}
 \left(\prod_{i=1}^d \alpha_i
 \right)\\
 &\times
 \left(\prod_{k=1}^d (\alpha^{-2}_k\rho_k-\frac{1}{2}\sum_{j=1}^d\alpha^{-2}_j\rho_j)^{1/p_k}\right)^{\frac{1}{\Sigma-1}},
 \end{split}
\end{equation}
where we express $\mathcal{L}$ as a function of the eigenvalues $\rho_k$ of $g^{-1}R$, and $\alpha_k$ can be determined through 
\begin{equation}
 \frac{\partial \mathcal{L}(\rho)}{\partial \rho_k}=\frac{1}{2\kappa \alpha_k^2}\prod_{i=1}^d \alpha_i.   
\end{equation}
The flow equation (\ref{flowL}) allows computing the small $\lambda$ expansion of $\mathcal{L}(g^{-1} R)$ (see the Supplemental Material).

Finally, in two dimensions, one can couple \eqref{se} to a Jackiw-Teitelboim-like gravity. We find that an undeformed matter theory coupled with a deformed Jackiw-Teitelboim-like gravity is dynamically equivalent to a deformed theory
coupled to a Jackiw-Teitelboim-like gravity.
The details are given in the Supplemental Material.

\section{ Including marginal flows in general deformations}
One can consider Ricci-based gravity theories associated with more general deformations. For instance, the stress tensor deformation originating from Eddington-inspired Born-Infeld gravity \cite{BeltranJimenez:2017doy} plays a role in $d=4$ $T\bar{T}$-like deformations of Abelian gauge theories \cite{Morone:2024ffm}.
For a stress tensor flow driven by an arbitrary operator $f(\tau_i)$ with parameter $\lambda$, the associated $B$ function satisfies the flow equation:
\begin{equation}\label{flowgenB2}
  \frac{\partial B}{\partial \lambda}  =
  f(\tau_i) \prod_{k=1}^d \alpha_k ,~~~\tau_i=
\alpha_i\partial_{\alpha_i}B\prod_{j=1}^d \alpha^{-1}_j.
\end{equation}
One can also include marginal flows by replacing $\alpha_i \rightarrow \alpha_i/\beta_i$ with $\prod_{k=1}^d \beta_k=1$ in the $B$ function.
The eigenvalues $\tau_i$ are modified as $\tau_i(\alpha_j )\rightarrow \tau_i (\alpha_j/\beta_j )$,
and the form of the flow equation (\ref{flowgenB2}) remains unchanged.
The flow equations associated with the $\beta$-parameters are,
\begin{equation}
 \beta_i \frac{\partial B}{\partial \beta_i} 
 -\beta_j \frac{\partial B}{\partial \beta_j}  =
 \alpha_i \frac{\partial B}{\partial \alpha_i} 
 +\alpha_j \frac{\partial B}{\partial \alpha_j} 
  =-(\tau_i-\tau_j)\prod_{k=1}^d \alpha_k .
\label{eqbeta}
\end{equation}
Therefore, it is possible to incorporate commutative marginal flows for any stress tensor deformation that admits a geometric realization.
It follows from (\ref{flowL}) that the associated Ricci-based gravity action should be modified as 
$\mathcal{L}(\rho_j )\rightarrow \mathcal{L} (\beta^{-2}_j\rho_j)$. 
\section{Conclusions}
This work introduces a geometric formulation for the combination of  $T\bar{T}$ and root-$T\bar{T}$ deformations in $d=2$. We demonstrate that these deformations can be classically formulated by coupling the undeformed theory with a massive gravity action. Additionally, we extend the geometric framework to encompass various stress-energy tensor deformations in higher dimensions. These deformations are related to several well-known theories, including ModMax and its Born-Infeld-like extension.  Furthermore, we study the Ricci-based gravities associated with such deformations. These findings might have broad implications in key areas of string theory and holography, improving our understanding of the effects of stress tensor deformations. Note also that our approach appears suitable for studying various irrelevant and marginal deformations. However, challenges arise in finding exact solutions for more complex 
$B$ functions in (\ref{eqB}). Not all deformations will lead to explicit or unique solutions for the relevant constraints, which generalize (\ref{flowgenB2}) and (\ref{eqbeta}).

There are several compelling avenues for future exploration stemming from our current work. A natural question is whether our formulation allows for the study of root-$T\bar{T}$ or more general deformations at the quantum level. The quantization of root-$T\bar{T}$ deformed theory poses a complex challenge, although some relevant progress has been made recently \cite{Ebert:2024zwv, Hadasz:2024pew}. We anticipate that our formulation could offer insights into this intricate issue. 
Another avenue worth exploring is investigating the holographic dictionary of these deformations.
Further, exploring the corresponding realization in celestial holography \cite{Pasterski:2016qvg} would be valuable, as proposed in \cite{He:2022zcf}, which offers a potential avenue for constructing UV-complete gravity theories. The link between stress-energy flows and classical string or D-brane actions can provide insights into the UV completeness of deformed theories. Consequently, one can envisage constructing counterparts \cite{He:2022zcf, He:2023lvk} in the framework of celestial holography to investigate their role in UV-complete theories. 

\textit{Note added:}
After our work was submitted to arXiv, \cite{Tsolakidis:2024wut} appeared, also investigating the massive gravity description of the root-$T\bar{T}$ deformation and finding results consistent with ours. \cite{Tsolakidis:2024wut} also examined deformations with explicit $\lambda$-dependence across various dimensions.
In addition, an auxiliary field method to define integrable deformations of the principal chiral model was discussed in \cite{Ferko:2024ali, Bielli:2024khq, Bielli:2024ach}. Exploring the potential connections between the two approaches remains an intriguing open problem.

\begin{acknowledgments}

\section{Acknowledgments}
The authors thank Dmitri Sorokin for his valuable discussions and comments. H. B-A. expresses his gratitude to the Department of Physics and Astronomy "Galileo Galilei" at the University of Padova for their generous hospitality during the concluding phase of this project.  S. H. acknowledges financial support from the Max Planck Partner Group, the Fundamental Research Funds for the Central Universities, and the Natural Science Foundation of China Grants No. 12075101, No. 12235016, No. 12347209, and No. 12475053. H. O. is supported by the National Natural Science Foundation of China, Grant No. 12205115, and by the Science and Technology Development Plan Project of Jilin Province of China, Grant No. 20240101326JC. T. M. and R. T. received partial support from the INFN project SFT and the the Prin (Progetti di rilevante interesse nazionale)  Project No. 2022ABPBEY, with the title ``Understanding quantum field theory through its deformations'', funded by the Italian Ministry of University and Research.

\end{acknowledgments}



 \onecolumngrid

 \appendix
 \section{Appendix}

\section{Metric formulation}
This section presents the geometric formulation of the combination of  $T\bar{T}$ and root-$T\bar{T}$ deformation in the metric approach. One can choose a gauge such that
\begin{equation}
	e^{-1}f=\sqrt{g^{-1}h}.
\end{equation}
The gravity action $S_{\mathrm{grav}}$ can be expressed in terms of the metrics $g_{\mu\nu}$ and $h_{\mu\nu}$ by using the relation
\begin{equation}
	\det\left(\sqrt{g^{-1}h}\right)=\sqrt{(\det g)^{-1}\det(h)}=\frac12\left[\operatorname{tr}(\sqrt{g^{-1}h})\right]^2-\frac12\operatorname{tr}(g^{-1}h).
\end{equation}
We have
\begin{align}
	y_1&=\mathrm{tr}(\sqrt{g^{-1}h})=\sqrt{z_1+ \sqrt{2 z_1^2-2z_2}},\\
	y_2&=\mathrm{tr}(g^{-1}h)=z_1,
\end{align}
where we denote $z_n=\mathrm{tr}[(g^{-1}h)^n]$. Then the gravity action $S_{\mathrm{grav}}$ can be written as
\begin{equation}\label{sz}
	\begin{split}
		S_{\mathrm{grav}}[g,h]
		=& \frac{1}{2\lambda}\int d^2x \sqrt{\det g}
		\Big(2 \sqrt{z_1-\sqrt{2} \sqrt{z_1^2-z_2}} \sinh \frac{\gamma }{2}\\&-2 \sqrt{\sqrt{2} \sqrt{z_1^2-z_2}+z_1} \cosh \frac{\gamma }{2}+\sqrt{2} \sqrt{z_1^2-z_2}+2
		\Big).  
	\end{split}
\end{equation}
One can also verify the flow equations in the metric formulation. The stress-energy tensor and related quantities are
\begin{align}
	T^{\mu}_{~\nu}=& 2 \frac{1}{\sqrt{\det h}}\frac{\delta S_{\gamma,\lambda}}{\delta h_{\mu\alpha}}h_{\alpha\nu}= 2 \frac{1}{\sqrt{\det h}}\frac{\delta S_{\mathrm{grav}}}{\delta h_{\mu\alpha}}h_{\alpha\nu},\\
	T^{\nu}_{~\nu}=&\frac{\sqrt{2} \sqrt{z_1-\sqrt{2} \sqrt{z_1^2-z_2}} \sinh \frac{\gamma }{2}}{\lambda\sqrt{z_1^2-z_2}}-\frac{\sqrt{2} \sqrt{\sqrt{2} \sqrt{z_1^2-z_2}+z_1} \cosh \frac{\gamma }{2}}{\lambda\sqrt{z_1^2-z_2}}+\frac{2}{\lambda},\\ T^{\mu}_{~\nu}T^{\nu}_{~\mu}=&\frac{2 \sqrt{2} \sqrt{z_1-\sqrt{2} \sqrt{z_1^2-z_2}} \sinh \frac{\gamma }{2}}{\lambda ^2 \sqrt{z_1^2-z_2}}-\frac{2 \sqrt{2} \sqrt{\sqrt{2} \sqrt{z_1^2-z_2}+z_1} \cosh \frac{\gamma }{2}}{\lambda ^2 \sqrt{z_1^2-z_2}}
	\\&+\frac{-2 \left(\sqrt{2 z_2-z_1^2} \sinh \gamma +z_1 \left(\cosh \gamma+z_1\right)-z_2\right)}{\lambda ^2 \left(z_1^2-z_2\right)}
	.
\end{align}
The action satisfies the flow equations
\begin{align}
	\frac{\partial S_{\gamma,\lambda}}{\partial \lambda}&=\int d^2x \sqrt{\det h}
	\left(\frac{1}{2}T^\mu_\nu T^\nu_\mu-\frac{1}{2}(T^\mu_\mu)^2\right)
	,\\
	\frac{\partial S_{\gamma,\lambda}}{\partial \gamma}&=\int d^2x \sqrt{\det h}
	\sqrt{\frac{1}{2}T^\mu_\nu T^\nu_\mu-\frac{1}{4}(T^\mu_\mu)^2}.
\end{align}
\section{Holographic boundary conditions}
When $e^a_\mu$ is on-shell, one can express $\alpha_j$ and $\tau_j$ in terms of $\tau^{[0]}_k$ as:
\begin{align}  \alpha_j&=\beta_j\left(\lambda(\tau^{[0]}_j)^{-1}\Big(\prod_{k=1}^d \tau^{[0]}_k{}^{1/p_k}\Big)^{\frac{1}{\Sigma-1}}+1\right)^{1/p_j},  \\
	\tau_j&=\left(\lambda\Big(\prod_{k=1}^d \tau^{[0]}_k{}^{1/p_k}\Big)^{\frac{1}{\Sigma-1}}+\tau^{[0]}_j\right)\prod_{k=1}^d\alpha^{-1}_k.
\end{align}
To derive the root-$T\bar T$ deformed boundary conditions, we take $d=2$, $p_k=1$, $\beta_1=e^{\frac{\gamma}{2}}$, $\beta_2=e^{-\frac{\gamma}{2}}$ and $\lambda\rightarrow 0$. 
We get, $\alpha_j=\beta_j$ and $\tau_j=\tau^{[0]}_j$. We write $T^{[0]}{}^\mu_{~\nu}$ explicitly:
\begin{equation}
	T^{[0]}{}^\mu_{~\nu} =  \left(\begin{array}{cc}a & b \\ c & d\end{array}\right)\,,
\end{equation}
which can be diagonalized by
\begin{equation}
	U=\left(
	\begin{array}{cc}
		\sqrt{a^2-2 a d+4 b c+d^2}+a-d \,\,\,&\,\,\, -\sqrt{a^2-2 a d+4 b c+d^2}+a-d \\
		2 c \,\,\,&\,\,\, 2 c \\
	\end{array}
	\right).
\end{equation}
Since $U$ also diagonalises $g^{-1}h$, we get
\begin{align}  
	g^{-1}h&= U \mathrm{diag}(e^{\gamma},e^{-\gamma})U^{-1}=\left(
	\begin{array}{cc}
		\frac{(a-d) \sinh (\gamma )}{\sqrt{(a-d)^2+4 b c}}+\cosh (\gamma ) & \frac{2 b \sinh (\gamma )}{\sqrt{(a-d)^2+4 b c}} \\
		\frac{2 c \sinh (\gamma )}{\sqrt{(a-d)^2+4 b c}} & \frac{(d-a) \sinh (\gamma )}{\sqrt{(a-d)^2+4 b c}}+\cosh (\gamma ) \\
	\end{array}
	\right),\\
	\Rightarrow~~
	h_{\mu\nu}& = g_{\mu\nu}\cosh{\gamma} + \frac{\sinh{\gamma}}{ \sqrt{\frac{1}{2}T^{[0]}{}^\mu_{~\nu} T^{[0]}{}^\nu_{~\mu}-\frac{1}{4}(T^{[0]}{}^\mu_{~\mu})^2}} \tilde{T}^{[0]}_{\mu\nu}\,,
\end{align}
where $\tilde{T}^{[0]}$ is the traceless part of $T^{[0]}$. The deformed boundary conditions for the stress-energy tensors are $T^\mu_{~\nu}=T^{[0]}{}^\mu_{~\nu}$.
\section{On-shell geometric formulation of MMBI and generalized Nambu-Goto}
When the stress-energy tensor of the undeformed theory has two degenerate eigenvalues $\tau^{[0]}_1$ of multiplicity $d_1$ and $\tau^{[0]}_2$ of multiplicity $d_2$, we propose an alternative gravity action:
\begin{equation}\label{degenerate}
	\begin{split}
		S_{\mathrm{grav}} =\frac{1}{\lambda^{\Sigma-1}} 
		\int d^d x \left[\left(\frac{e^{\frac{\gamma }{d_2}} \left(d_2 y_1\mp\sqrt{d_1 d_2 \left(\left(d_1+d_2\right) y_2-y_1^2\right)}\right)}{d_2 \left(d_1+d_2\right)}\right)^{p_2}-1\right]^{\frac{d_2}{p_2}} \\
		\times
		\left[\left(\frac{e^{-\frac{\gamma }{d_1}} \left(d_1 y_1\pm\sqrt{d_1 d_2 \left(\left(d_1+d_2\right) y_2-y_1^2\right)}\right)}{d_1 \left(d_1+d_2\right)}\right)^{p_1}-1\right]^{\frac{d_1}{p_1}} .   
	\end{split}
\end{equation}
When $e^a_\mu$ is on-shell, the eigenvalues $\alpha_k$ and $\tau_k$ have the same degenerate structure as $\tau^{[0]}_k$. One finds
\begin{align}
	\frac{\partial S_{\mathrm{grav}}}{\partial \lambda}
	&=-(\Sigma-1)\int d^d x \det f (\tau_1^{d_1/p_1}\tau_2^{d_2/p_2})^{\frac{1}{\Sigma-1}},\\
	\frac{\partial S_{\mathrm{grav}}}{\partial \gamma}
	&=\int d^d x \det f (\tau_2-\tau_1).
\end{align}
For the MMBI case, we take $d_k=p_k=2$ and \eqref{degenerate} becomes
\begin{equation}\label{explmmbi}
	\begin{split}
		S_{\mathrm{grav}} =& \frac{1}{\lambda}\int d^4 x \det e \left[1-\frac{y_2}{2} \cosh{ \gamma}\pm\frac{y_1 }{4}  \sqrt{4 y_2-y_1^2}\sinh{ \gamma}+\left({\frac{y_1^2}{8}-\frac{y_2}{4} }\right)^{\!\!2}\,\right] .  
	\end{split}
\end{equation}
For the generalized Nambu-Goto case, we take $d_1=p_1=1$, $d_2=p_2=d-1$, and \eqref{degenerate} becomes
\begin{equation}
	\begin{split}
		S_\mathrm{grav} =& \frac{1}{\lambda}\int d^dx \det e\,\left[\left(\frac{(d-1) y_1\mp\sqrt{(d-1) \left(d y_2-y_1^2\right)}}{(d-1) d}\right)^{d-1}-1\right] \left[\frac{\pm\sqrt{(d-1) \left(d y_2-y_1^2\right)}+y_1}{d}-1\right]. 
	\end{split}
\end{equation}
\section{Flow equations of $\mathcal{L}(g^{-1} R)$}
In this section, we derive flow equations of $\mathcal{L}(g^{-1} R)$, which enable us to compute the perturbative expansion of $\mathcal{L}(g^{-1} R)$ in small deformation parameters.
When total action is extremized with respect to $h_{\mu\nu}$, only explicit dependencies must be considered when differentiating the Lagrangian with respect to fields or parameters. We have
\begin{align}
	\frac{\partial}{\partial R_{\mu\nu}} \left(\sqrt{\det g} \mathcal{L}(g^{-1} R)\right)&= \frac{\partial}{\partial R_{\mu\nu}}\left( \frac{1}{2\kappa} \sqrt{\det h} h^{\alpha\beta}R_{\alpha\beta}\right)\label{dR},\\
	\frac{\partial}{\partial \lambda} \left(\sqrt{\det g} \mathcal{L}(g^{-1} R)\right) &= \frac{\partial}{\partial \lambda}\left( \sqrt{\det g} B(g^{-1}h)\right).\label{dl}
\end{align}
It follows from the equations of motion of $h_{\mu\nu}$ that
\begin{equation}\label{einstein}
	g^{\mu\alpha}R_{\alpha\nu}=
	\kappa g^{\mu\alpha}h_{\alpha\beta}
	T^\beta_{~\nu}
	-\frac{\kappa}{d-2}g^{\mu\alpha}h_{\alpha\nu}T^\beta_{~\beta},
\end{equation}
and therefore $g^{\mu\alpha}R_{\alpha\nu}$ and $g^{\mu\alpha}h_{\alpha\nu}$ can be diagonalized simultaneously.
Denoting $\rho_k$ as the eigenvalues of $g^{\mu\alpha}R_{\alpha\nu}$ and writing $\mathcal{L}$ as a function of $\rho_k$,  equation (\ref{einstein}) can be written as
\begin{equation}\label{eigeneinstein}
	\kappa\tau_k=\alpha^{-2}_k\rho_k-\frac{1}{2}\sum_{j=1}^d\alpha^{-2}_j\rho_j,
\end{equation}
and
equation (\ref{dR}) leads to
\begin{equation}\label{alpharho}
	\frac{\partial \mathcal{L}(\rho)}{\partial \rho_k}=\frac{1}{2\kappa \alpha_k^2}\prod_{i=1}^d \alpha_i~~\Rightarrow \alpha_k=(2 \kappa )^{\frac{1}{d-2}}  \left(\frac{\partial \mathcal{L}(\rho)}{\partial \rho_k}\right)^{-1}
	\left(
	\prod_{i=1}^d  \frac{\partial \mathcal{L}(\rho)}{\partial \rho_i}
	\right)^{\frac{1}{2d-4}}.
\end{equation}
Using (\ref{flowgen}), (\ref{dl}) and (\ref{eigeneinstein}), we get the flow equation of $\mathcal{L}(\rho)$ with respect to $\lambda$:
\begin{equation}
	\frac{\partial \mathcal{L}(\rho)}{\partial \lambda}=-(\Sigma-1)\kappa^{-\frac{\Sigma}{\Sigma-1}}
	\left(\prod_{i=1}^d \alpha_i
	\right)
	\left(\prod_{k=1}^d (\alpha^{-2}_k\rho_k-\frac{1}{2}\sum_{j=1}^d\alpha^{-2}_j\rho_j)^{1/p_k}\right)^{\frac{1}{\Sigma-1}},
\end{equation}
where one should also substitute (\ref{alpharho}) into the right-hand site. The initial condition can be obtained using the limit $\lambda\rightarrow 0$. We have: 
\begin{align}
	\alpha_k&=\beta_k+O(\lambda)
	\\
	\mathcal{L}(\rho) &=\frac{1}{2\kappa}  \sum_{k=1}^d \beta^{-2}_k \rho_k+O(\lambda).  
\end{align}
Using the flow equation and the initial condition, we can recursively solve the $\lambda$ expansion of $\mathcal{L}(\rho)$. Up to order $\lambda$ we find
\begin{equation}
	\mathcal{L}(\rho) =\frac{1}{2\kappa}  \sum_{k=1}^d \beta^{-2}_k \rho_k
	-\lambda(\Sigma-1)
	\kappa^{-\frac{\Sigma}{\Sigma-1}}
	\left(\prod_{k=1}^d (\beta^{-2}_k\rho_k-\frac{1}{2}\sum_{j=1}^d\beta^{-2}_j\rho_j)^{1/p_k}\right)^{\frac{1}{\Sigma-1}}
	+O(\lambda^2).   
\end{equation}
For pure $( \det T)^{\frac{1}{d-p}}$ deformations with $\beta_k=1$ and $p_k=p$, the expression reduces to
\begin{equation}
	\begin{split}
		\mathcal{L} =&\frac{1}{2\kappa} \sum_{k=1}^d  \rho_k
		-\lambda(d/p-1)
		\kappa^{-\frac{d}{d-p}} 
		\prod_{k=1}^d (\rho_k-\frac{1}{2}\sum_{j=1}^d\rho_j)^{\frac{1}{d-p}}
		+O(\lambda^2)\\=&
		\frac{1}{2\kappa} \tr(g^{-1}R)
		-\lambda(d/p-1)
		\kappa^{-\frac{d}{d-p}} 
		\det (g^{-1}R-\frac{1}{2}\tr(g^{-1}R))^{\frac{1}{d-p}}
		+O(\lambda^2).    
	\end{split}  
\end{equation}
However, for more general deformations, it is difficult to express the eigenvalues $\rho_k$ in terms of $\tr[(g^{-1}R)^k]$ explicitly.
\section{Coupling to flat Jackiw-Teitelboim-like  gravity 
	action in two dimensions}
In two dimensions, we couple the action (\ref{se}) to a flat space Jackiw-Teitelboim-like gravity action in the first-order formalism for the zweibein $f^a_\mu$ and a vacuum energy term:
\begin{align}
	S=
	S_{JT}
	-\Lambda\int d^2 x \det f 
	+\int d^2 x \det e B(e^{-1}f)+
	S_0[\phi,e^a_\mu], \label{Stotal2d}\\
	S_{JT}=\frac{1}{\kappa}\int d^2 x
	\epsilon^{\alpha\beta}
	(\epsilon_{ac}\sigma^c(\partial_\alpha f^a_{\beta}-\epsilon^a_{~b}\omega_\alpha f^b_{\beta})+\varphi\partial_\alpha\omega_\beta).
\end{align}
The equation of motion for $f$ gives
\begin{equation}
	\frac{1}{\kappa}
	\epsilon^{\alpha\beta}\epsilon_{ab} u^b_\beta-\Lambda\epsilon^{\alpha\beta}\epsilon_{ab} f^b_\beta+\det e\frac{\partial B}{\partial f^a_\alpha}=0,
\end{equation}
where we defined $u^a_\alpha=\partial_\alpha \sigma^a-\epsilon^a_{~c}\sigma^c\omega_\alpha$. The solution is
\begin{equation}
	f^*= \frac{\sqrt{2 w_2-w_1^2} w_1 \sinh \frac{\gamma }{2}+\left(w_1^2-2 w_2\right) \cosh \frac{\gamma }{2}}{\left(w_1^2-2 w_2\right) (1-\lambda  \Lambda )}e-\frac{\lambda -\frac{2 \kappa  \sinh \frac{\gamma }{2}}{\sqrt{2 w_2-w_1^2}}}{\kappa -\kappa  \lambda  \Lambda }   u,
\end{equation}
where $w_n=\mathrm{tr}[(e^{-1}u)^n]$. In terms of $y_n$, we find
\begin{equation}
	y^*_1= \frac{2 \kappa  \cosh \frac{\gamma }{2}-\lambda  w_1}{\kappa -\kappa  \lambda  \Lambda },~~~y_2^*= \frac{2 \kappa ^2 \cosh \gamma +\lambda  \left(-2 \kappa  \sqrt{2 w_2-w_1^2} \sinh \frac{\gamma }{2}-2 \kappa  w_1 \cosh \frac{\gamma }{2}+\lambda  w_2\right)}{\kappa ^2 (\lambda  \Lambda -1)^2}.
\end{equation}
Substituting back into  (\ref{Stotal2d}), we get
\begin{equation}
	\begin{split}
		S=&
		\int d^2x \det e
		\left(\frac{2 \kappa ^2 \lambda  \Lambda +4 \kappa ^2 \cosh \gamma -4 \kappa ^2+2 \kappa  \lambda  \sqrt{2 w_2-w_1^2} \sinh \frac{\gamma }{2}-2 \kappa  \lambda  w_1 \cosh \frac{\gamma }{2}+\lambda ^2 w_1^2-\lambda ^2 w_2}{2 \kappa ^2 \lambda  (\lambda  \Lambda -1)}
		\right)\\ &+\int d^2x \epsilon^{\alpha\beta}\varphi\partial_\alpha\omega_\beta+S_0.   
	\end{split}
\end{equation}
Denoting the eigenvalues of $e^{-1}u$ as $\nu_k$, the action can be simplified as
\begin{equation}
	\begin{split}
		S=&
		\int d^2x \det e
		\left(\frac{\kappa ^2 \lambda  \Lambda +\kappa ^2 e^{-\gamma }+\kappa ^2 e^{\gamma }-2 \kappa ^2-\kappa  \lambda  e^{\gamma /2} \nu _1-\kappa  \lambda  e^{-\frac{\gamma }{2}} \nu _2+\lambda ^2 \nu _1 \nu _2}{\kappa ^2 \lambda  (\lambda  \Lambda -1)}
		\right)\\&+\int d^2x \epsilon^{\alpha\beta}\varphi\partial_\alpha\omega_\beta+S_0,  
	\end{split}
\end{equation}
which can be interpreted as a matter theory $S_0$
coupled to a deformed Jackiw-Teitelboim-like gravity. 
Alternatively, integrating out the vielbein $e^a_\mu$ in the action (\ref{Stotal2d}) results in a matter theory that is deformed by both $T\bar{T}$ and root-$T\bar{T}$, coupled to Jackiw-Teitelboim-like gravity. Consequently, the dynamics of a matter theory $S_0$ coupled to a deformed Jackiw-Teitelboim-like gravity is equivalent to that of a matter theory subjected to $T\bar{T}$ and root-$T\bar{T}$ deformations coupled to Jackiw-Teitelboim-like gravity.

\end{document}